# ProHap Explorer: Visualizing Haplotypes in Proteogenomic Datasets

*Jakub Vašíček[1,2,$], Dafni Skiadopoulou[1,2], Ksenia G. Kuznetsova[1,2,3], Lukas Käll[4,*], Marc Vaudel[1,2,5,*], Stefan Bruckner[6,*,$]*

[1] Mohn Center for Diabetes Precision Medicine,
  Department of Clinical Science,
  University of Bergen, Bergen, Norway

[2] Computational Biology Unit,
  Department of Informatics,
  University of Bergen, Bergen, Norway

[3] Metabolism Program,
  Broad Institute of MIT and Harvard, Cambridge, MA, USA

[4] Science for Life Laboratory,
  Department of Gene Technology,
  KTH - Royal Institute of Technology, Stockholm, Sweden
  orcid: 0000-0001-5689-9797

[5] Department of Genetics and Bioinformatics, Health Data and Digitalization,
  Norwegian Institute of Public Health, Oslo, Norway

[6] Chair of Visual Analytics,
  Institute for Visual and Analytic Computing,
  University of Rostock, Rostock, Germany

[*] These authors jointly supervised the work
[$] To whom correspondence should be addressed

Corresponding authors:

Jakub Vašíček - jakub.vasicek@uib.no;
Stefan Bruckner - stefan.bruckner@uni-rostock.de

## Abstract

In mass spectrometry-based proteomics, experts usually project data onto a single set of reference sequences, overlooking the influence of common haplotypes (combinations of genetic variants inherited together from a parent). We recently introduced *ProHap*, a tool for generating customized protein haplotype databases. Here, we present *ProHap Explorer*, a visualization interface designed to investigate the influence of common haplotypes on the human proteome. It enables users to explore haplotypes, their effects on protein sequences, and the identification of non-canonical peptides in public mass spectrometry datasets. The design builds on well-established representations in biological sequence analysis, ensuring familiarity for domain experts while integrating novel interactive elements tailored to proteogenomic data exploration. User interviews with proteomics experts confirmed the tool's utility, highlighting its ability to reveal whether haplotypes affect proteins of interest. By facilitating the intuitive exploration of proteogenomic variation, *ProHap Explorer* supports research in personalized medicine and the development of targeted therapies.

# 1. Introduction

A key concept in human genetics is haplotypes, which are sets of alleles inherited together from a parent. The increasing availability of population-wide genomic data sets has enabled the mapping of haplotypes across populations, facilitating genotype imputation and genome-wide association studies (GWAS) [1]. While genomics provides valuable insights into the genetic blueprint, it alone is insufficient to capture the complexity of biological systems.

Proteomics, the large-scale study of proteins, offers a deeper understanding of how genes are expressed and function within the body. In proteomics, data is usually projected onto a single reference proteome, which disregards the genetic diversity present in human populations. The field of proteogenomics has made efforts to allow the inspection of variant sequences in proteomics [2], [3].

Protein haplotypes, which are unique protein sequences encoded by sets of alleles in linkage disequilibrium (LD), can be discovered through proteomics [4]. Accounting for these protein haplotypes will enable us to fine-chart the human proteome, thereby acknowledging the diversity across populations. This approach not only enhances our understanding of proteins and their function and regulation, but also has the potential to improve the accuracy of biomarker discovery and therapeutic interventions.

Proteogenomic datasets are inherently multimodal, encompassing information concerning genetic variation, mRNA splicing, encoded protein sequences, peptide identifications, and various technical properties and quality metrics. Interpreting these datasets is challenging because experts need to maintain links between these dimensions. For instance, determining which regions of genes encode identified peptides and whether they cover important variant loci or splicing junctions is difficult without a clear visual overview. A robust visualization platform can assist experts in navigating these complex datasets, enabling them to extract relevant information in a meaningful and effective manner.

We have previously developed ProHap [5], a tool that creates databases of protein haplotypes from genotype panels of human populations, and allows experts to inspect the presence of these haplotypes in mass spectrometry datasets. Along with ProHap, we published a database of protein haplotypes derived from the genotypes of the 1000 Genomes Project [6], [7]. We have used this database to search a publicly available mass spectrometry dataset of healthy human tissues [8] to identify novel variant peptides, and explore their presence across tissues.

In this paper, we present *ProHap Explorer*, a web-based visual interface for proteogenomic data obtained using the *ProHap* workflow. It allows experts to browse the published database of protein haplotypes from the 1000 Genomes Project, and view whether peptides have been identified that cover interesting features in the protein sequences. The availability of such a platform will be instrumental in advancing our understanding of the connection between genetic variation and protein function, ultimately contributing to the field of personalized medicine and the development of targeted therapies.

## 2. Related Work

This work is rooted in several research areas, which we divide into three parts: proteogenomics, the field integrating genomic and mass spectrometry-based proteomic research; visualization of biological sequences; and visualization of proteomic data. This section reviews the relevant literature and existing tools in each of these domains, highlighting the contributions that have informed the development of *ProHap Explorer*.

### 2.1 Mass spectrometry-based proteogenomics

Mass spectrometry-based analysis is the reference platform for sequence-level proteomics [9]. In a so-called bottom-up proteomic experiment, proteins in biological samples are digested using an enzyme (e.g., trypsin) into peptides. These peptides are then separated using liquid chromatography based on their chemical properties before undergoing a tandem mass spectrometry (MS/MS) measurement. The obtained spectra are searched against a database of expected protein sequences using search engine software, producing a set of peptide-spectrum matches (PSMs) [10], [11], [12]. Rescoring methods are applied to estimate the false discovery rate and posterior error probabilities of PSMs [13], [14].

In proteogenomics, databases of protein sequences are obtained by aligning genetic variants with mRNA or cDNA sequences of spliced transcripts, which are then translated *in silico* to obtain the resulting protein sequences [3]. Tools such as *py-pgatk* [15] or *PrecisionProDB* [16] facilitate this process. Alternatively, the databases are created from the *in silico* translation of custom mRNA sequences using *customProDB* [17]. Spooner et al. have developed *Haplosaurus* to inspect different haplotypes of specific proteins using a dataset of phased genotypes [18]. Building upon this work, we have introduced *ProHap* - a Python-based pipeline to create databases of protein haplotype sequences from panels of phased genotypes [5].

## 2.2 Biological sequence visualization

In the realm of genomic and transcriptomic visualization, several tools have been developed for the exploration and analysis of sequences. Genome browser tools such as the *UCSC Genome Browse*r [19] and *JBrowse* [20] provide robust platforms for visualizing genomic data, offering multiple tracks displaying various features aligned onto a common scale. Similarly, the *Integrative Genomics Viewer* (IGV) [21] is a popular tool that provides a high-performance, interactive platform for visualizing various types of genomic data, including sequence alignments and variant calls [22].

Ensembl [23] is a widely used resource that provides a comprehensive genome browser, integrating diverse genomic data such as gene annotations, sequence variants, and transcript splicing information. *gnomAD* [24], a key resource for human genetic variation, also offers a similar representation. Additionally, Vials is a tool that focuses on the visualization of alternative splicing events and their impact on protein-coding sequences [25].

Resources such as *UniProt* [26] and *neXtProt* [27] play a crucial role in the integration and interpretation of protein data, complementing genomic resources like Ensembl. *UniProt* is a generic, high-quality database of protein sequence and functional information, providing detailed annotations on protein structure, function, and interactions across all species. *neXtProt* extends this by offering a platform specifically designed for human proteins, integrating data from various sources to provide a complete and experimentally-validated representation of the human proteome. *UniProt* features viewer tools highlighting domains, post-translational modifications, annotated sequence features, and variant locations. *neXtProt* offers similar visualizations, including a viewer for annotated features, and for identifications of peptides across mass spectrometry datasets.

## 2.3 Visualization of proteomic data

A number of visualization approaches focus on other aspects of proteomic datasets beyond protein sequences. A large body of work has been done on visualizations of protein structures [28], which include visualizations of protein-protein interactions and protein complexes [29], and visualizations of important features within protein structures [30]. These structural visualizations are crucial for understanding the functional mechanisms of proteins and their interactions within the cellular environment.

Other tools provide visual representations of raw proteomic data for the purpose of quality assurance. Examples of such tools include *PeptideShaker* [31], *PDV* [32], and *Skyline* [33], which offer detailed visualizations to assess the quality and reliability of proteomic experiments and peptide identifications. These visualizations often include displaying the mass spectra, where individual peaks are annotated with the corresponding peptide fragments. Additionally, these tools often display the elution profile of a peptide - the quantity of the peptide eluted from the liquid chromatography column per retention time - shown as a line chart.

Large repositories of mass spectrometry data, such as *PRIDE* [34] and other members of the *ProteomeXchange* consortium [35], also offer visualizations of mass spectra annotated with corresponding fragment ions. These visualizations can be accessed for any of the billions of spectra available using the *Universal Spectrum Identifier* [36], providing a valuable resource for researchers to explore and validate their findings. Additionally, resources like *ProteomicsDB* [37] and *GPMdb* [38] offer comprehensive databases of proteomic data, including various visualization tools to facilitate data exploration and interpretation.

## 3. Data Modeling and Domain Goals

Despite significant advances in genome and proteome analysis tools, there remains a gap in integrating haplotypes, protein sequence variation, and peptide evidence from mass spectrometry into a single interactive visualization platform. Genome browsers like *Ensembl* and *IGV* provide rich annotations for genomic variants but lack direct integration with proteomic data. Proteogenomic pipelines can generate custom protein databases to detect variant peptides, yet they do not offer interactive exploration of how haplotypes shape the proteome. Proteomic resources such as *UniProt* provide overview of variation in protein sequences, but do not consider haplotypes, and do not display peptide-level evidence for variant peptides. *ProHap Explorer* addresses this need by enabling researchers to investigate how common haplotypes alter protein sequences and verify their presence in mass spectrometry datasets within an intuitive, interactive interface.

We identify four main layers of proteogenomic data (Figure 1), each representing a coordinate system to localize features of interest. The first layer is the gene (L1), defining the relevant region of the chromosome. Positions of features are encoded using their genomic coordinates -- the number of base pairs between the start of the chromosome and the feature locus. The second layer is the transcript (L2), which captures the splicing of genes and the location of start/stop codons, denoting the limits of the main open reading frame (mORF). Positions of features within the transcript are determined by the number of nucleotides from the start of the cDNA sequence. The translation of the mORF yields the protein (L3). In our context, a unique protein sequence encoded by a haplotype of a spliced mRNA is referred to as a *proteoform*. Feature positions in the protein are counted as the number of amino acid residues from the first methionine residue at the N-terminal end. Note that features located in the 5' untranslated region of the transcript will have a negative protein-level coordinate. Finally, the fourth dimension is the peptide (L4), representing a part of the protein that has been matched to a spectrum, the features that fall into the peptide range being localized relative to the position of the first amino acid of the peptide.

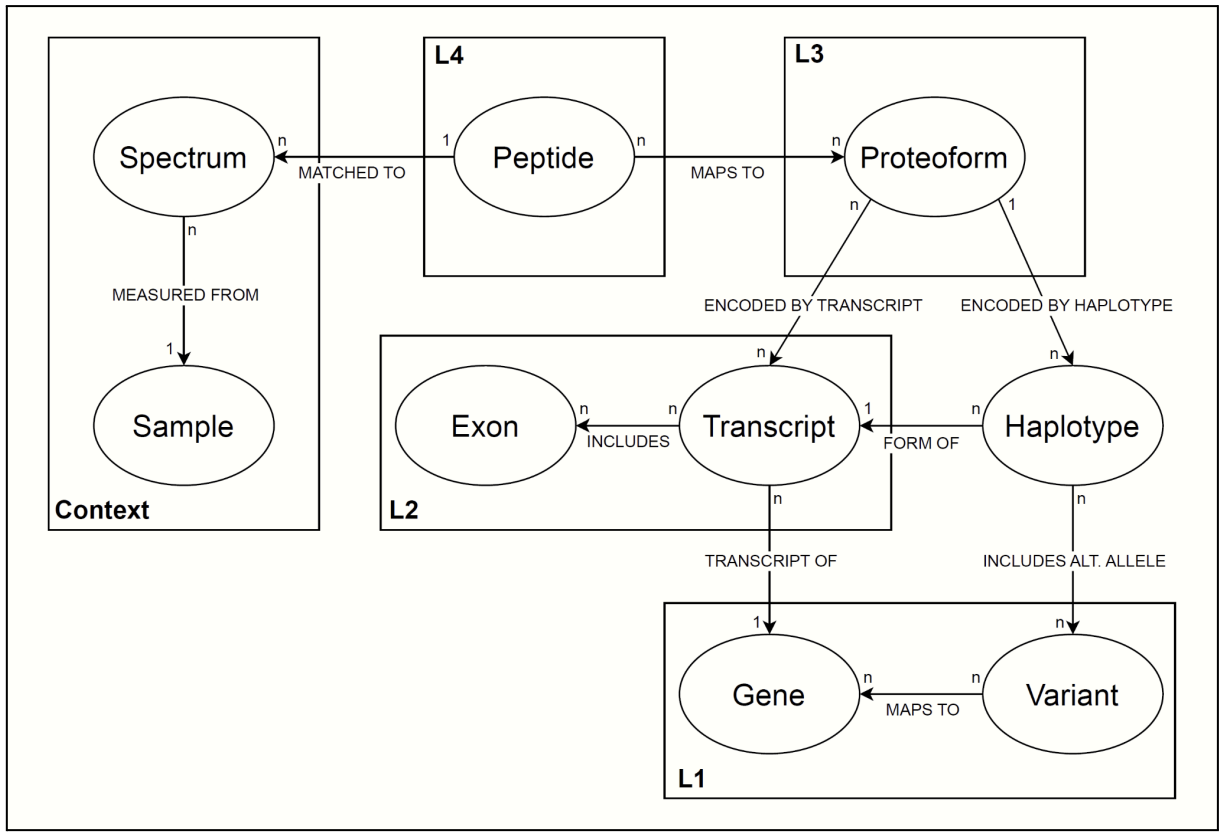

**Figure 1:** Relationships between the elements in proteogenomic datasets created using the ProHap workflow.

Peptides matched to mass spectra provide the interface between the theoretical proteome ("what can be encoded by genes") to the experimental data ("what has been observed in biological samples"). This context is essential for the expert to assign biological and medical relevance to regions of the proteome. However, inferring the presence of a particular proteoform from a set of peptide identifications, a problem known as protein inference [39], is often impossible to achieve with certainty. A single peptide sequence may be encoded by different genes, and almost always maps to multiple proteoforms encoded by the same gene. While statistical methods have been developed to tackle this problem [40], a visual approach will allow experts to gain an overview of the assignment of peptides to proteoforms.

Working in a multi-disciplinary team with expertise in bioinformatics, proteomics, and genomics, we have identified four key visualization tasks. These were clarified during the iterative development of ProHap and the adjacent workflows, and upon discussions with collaborators in meetings and conferences. The tasks can be summarized using the typology introduced by Munzner and Brehmer [41] as follows:

**T1** *Explore* unique protein sequences encoded by human genes and the identification rates of matching peptides (e.g., see which genes encode the highest number of variant peptides, and whether any of these are relevant to a particular research project).

**T2** *Locate* regions of a protein affected by variation (e.g., once a relevant gene has been identified, see which part of the protein sequence maps to variant peptides, and which splicing alternative has the best coverage).

**T3** *Lookup* confident peptide-spectrum matches for a protein of interest (e.g., once peptides of interest have been identified, review the quality of the peptide-spectrum matches).

**T4** *Browse* identifications of variant peptides across tissues and datasets (e.g., when investigating a particular variant, see in which tissues, or which mass spectrometry datasets these peptides have been identified).

### 3.1 Dataset Presented in ProHap Explorer

We have created a database of protein haplotypes using ProHap on the complete dataset of phased genotypes published by the 1000 Genomes Project [6]. Individual variants were thresholded at a 1% minor allele frequency, with no haplotype frequency threshold applied. Variants were aligned with spliced cDNA sequences according to Ensembl v.110 [42] before in-silico translation. Variants mapping to the untranslated regions of transcripts were retained in the haplotypes; however, only the translation of the main open reading frame (mORF) was used in the resulting database of protein haplotype sequences.

We used this sequence database to search a public mass spectrometry proteomic dataset of healthy human tissues [8], downloaded from the PRIDE repository PXD010154. For the search, we employed SearchGUI [43] version 4.3.1 and PeptideShaker [31] version 3.0.0, using the X!Tandem [11] and Tide [44] search engines. The highest scoring peptide-spectrum match (PSM) per spectrum was used for peptide identifications. The modification settings specified were carbamidomethylation of cysteine (Unimod #4) as a fixed modification, and oxidation of methionine (Unimod #35), deamidation of asparagine and glutamine (Unimod #7), and acetylation of the protein N-terminus as variable modifications. The maximum peptide length was set to 40 amino acids, with precursor and fragment ion tolerances set to 10 ppm and 0.05 Da, respectively.

Resulting PSMs were processed as described in [45] using Percolator [46] version 3.5, with features based on peptide retention time (DeepLC [47] version 1.1.2) and fragmentation predictors (MS2PIP [48] version 3.9.0), and filtered at a 1% estimated false discovery rate (FDR). We then used the ProHap Peptide Annotator pipeline to annotate the identified peptides with the corresponding transcripts, gene names, and matching alleles of genomic variants. All the data have been stored in a graph database, illustrated in Figure 1, for effective querying.

## 4. ProHap Explorer Interface

ProHap Explorer is a web-based visualization tool designed to facilitate the exploration and analysis of proteogenomic data by integrating haplotypic variation with peptide evidence from mass spectrometry datasets. Unlike existing tools that focus solely on genomic sequences or peptide identifications in isolation, ProHap Explorer introduces several novel capabilities, including: (1) linking haplotype-level genetic variation to observed peptide data, (2) an interactive dual-view interface for efficient exploration, (3) proteoform-aware peptide classification, and (4) dataset-specific filtering and highlighting features.

The interface consists of two main views: the *Explore View*, which displays a list of all human genes and allows users to discover which genes are best covered in the mass spectrometry datasets and most affected by variation, and the *Detail View*, which enables users to conduct an in-depth analysis of a gene of interest.

### 4.1 Explore view

The *Explore View* in *ProHap Explorer* (Figure 2) is designed to provide users with a comprehensive overview of human genes and their coverage in mass spectrometry datasets (T1). This view consists of two main components. The first component is a table that lists all human genes along with several key metrics: the number of unique proteoforms they encode, the total number of peptides matched to spectra in the mass spectrometry data, and the number of those peptides that are variant peptides. Users can sort the table by any of these variables, allowing for easy identification of genes with specific characteristics or levels of coverage. Clicking on a gene identifier will open the *Detail View* of this gene, as described below in Section 4.2.

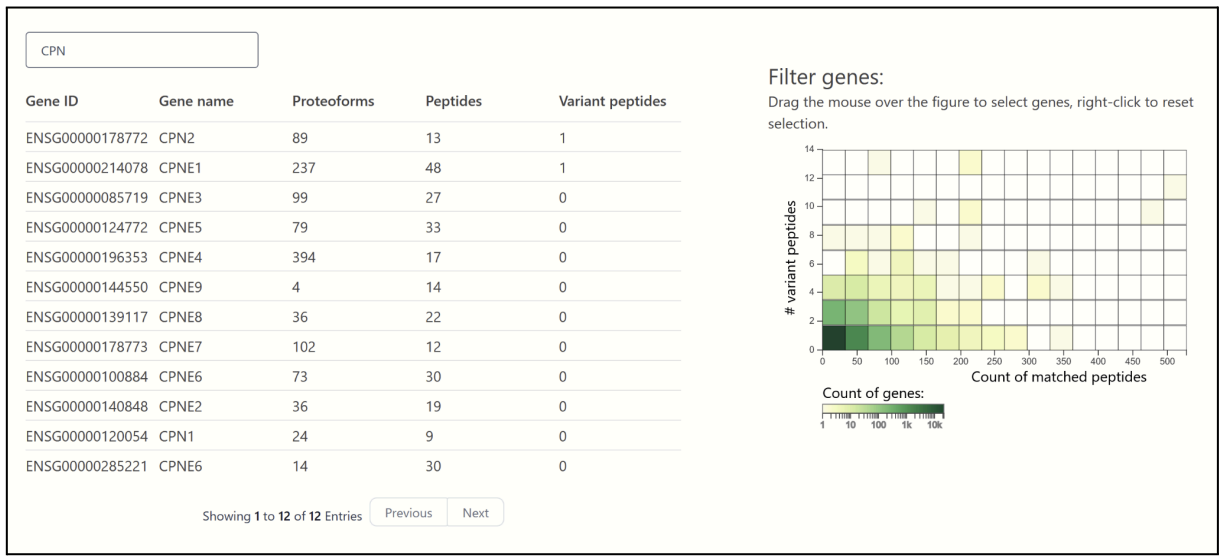

**Figure 2**: Explore view, searching for gene names containing the "CPN" substring.

In developing the *Explore View*, we considered several alternative visualizations in addition to the data table. One option was to visualize all the genes using a space-filling curve, with color-coded regions representing the number of identified peptides or identified variant peptides. Space-filling curves, such as Hilbert curves, can effectively display large datasets while preserving spatial relationships [49]. However, this approach was ultimately discarded because it can be challenging for users to interpret and navigate, especially when trying to quickly filter genes with multiple variant peptides or a high number of peptides overall. The complexity of space-filling curves can make it difficult to extract specific insights at a glance.

The chosen design of the second component is an interactive 2D histogram. This histogram visualizes the number of peptides matched to spectra on the x-axis and the number of variant peptides of those matched on the y-axis. Each cell in the heatmap is color-coded to represent the number of genes with the respective properties on a logarithmic scale. Users can interact with the histogram by clicking on a specific cell to select it or by making a rectangle selection by dragging the mouse over the heatmap. This selection will filter the genes shown in the table, enabling users to quickly focus on genes that meet specific criteria.

### 4.2 Detail view

Upon selection of a gene in the *Explore View*, the *Detail View* is displayed, which consists of two sections. Each section includes a primary visualization, a data table, and options to download raw data, providing a comprehensive and detailed analysis of the selected gene.

### 4.2.1 Splicing and Variation Section

The first section (Figure 3) represents the layers of genes and transcripts (L1, L2). The primary visualization in this section shows all the different splicing alternatives of the selected gene, with each row representing one transcript. Exons are aligned from the 5'-most to the 3'-most end and shown as rectangles, while introns are displayed as abbreviated lines. The locations of the canonical start and stop codons are highlighted, with the region in between, known as the main open reading frame (mORF), shown in a darker shade to distinguish it from untranslated regions (UTRs).

Vertical lines across all rows represent variant loci, with the variant type (single nucleotide polymorphism (SNP), in-frame insertion or deletion, and frameshift) encoded by different colors. This allows users to quickly identify the types and locations of genetic variants present in the coding regions of the gene (T2). Additionally, regions of exons that encode a sequence of a peptide matched in one of the mass spectrometry datasets are highlighted in dark blue, indicating parts of the protein that have been experimentally validated.

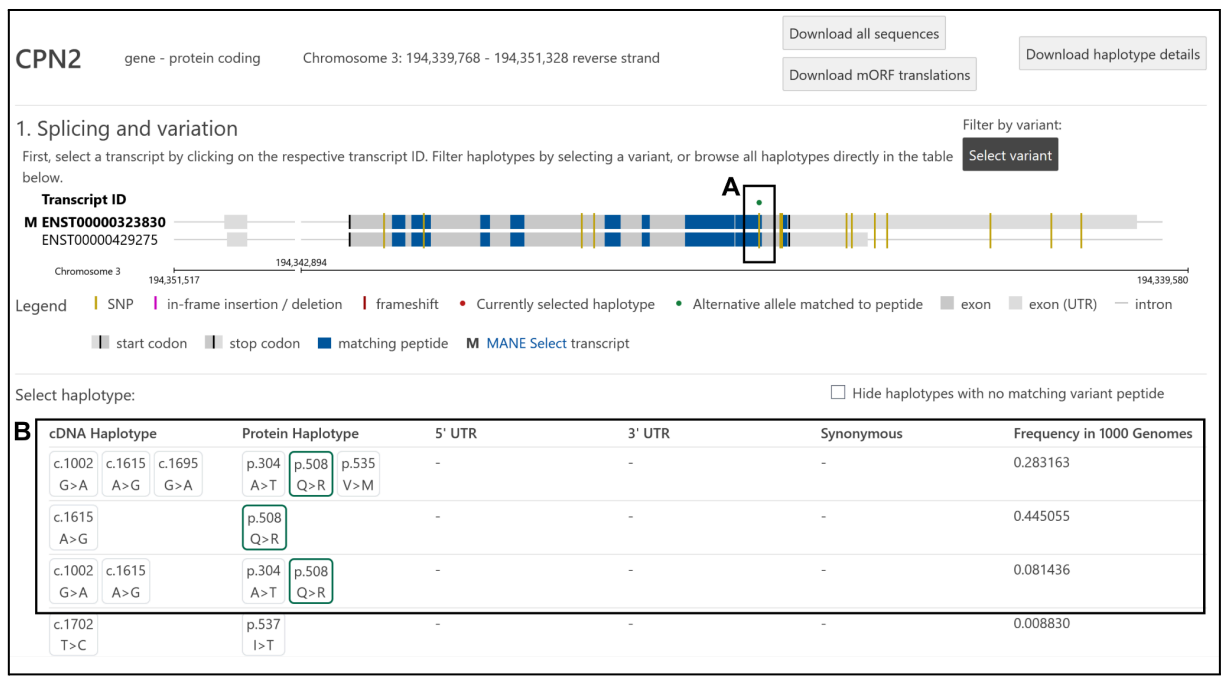

**Figure 3:** First section of the *Detail View*, showing the *CPN2* gene after the selection of a transcript, and before the selection of a haplotype. A: One variant encodes an amino acid substitution that has been observed in a peptide. B: This variant appears in three unique protein haplotype sequences, either in combination with one or two other substitutions, or individually.

This linear representation, with exons depicted as rectangles and introns as horizontal lines, is the most common way to represent splicing of genes. It is widely used in genome browsers, such as the *UCSC Genome Browser* [19] and *Ensembl* [23], making it a familiar format for experts. This familiarity allows them to interpret the figure much faster than other designs, enabling us to encode additional information such as variant loci and peptide mappings without overwhelming the user.

The data table in the first section becomes available upon the selection of a transcript. Rows in this table represent the unique non-canonical sequences possibly encoded by haplotypes in this transcript. For each of these proteoforms, the table shows the changes in the cDNA and protein sequence, and the observed frequency of the haplotype in the 1000 Genomes dataset. Amino acid changes that have been observed in an identified peptide are highlighted by a green outline, making it easy to estimate which haplotype have been observed in the analyzed samples.

*Interactions*: By clicking on the transcript identifier in the visualization of splicing alternatives, the user will bring up a data table of corresponding proteoforms. The user can then filter the displayed proteoforms by hiding haplotypes that do not encode any matched variant peptides using a checkbox above the table, and by restricting the view to haplotypes that include a variant of interest. The variant can be selected by clicking on the corresponding line element or using a searchable drop-down menu.

The user selects a proteoform by clicking on one of the table rows. Upon selection, the table row expands to reveal the different haplotypes possibly encoding this proteoform sequence, showing associated combinations of variants in the untranslated regions, and associated combinations of synonymous variants (i.e., not encoding an amino acid change) in the protein-coding regions. This iterative filtering approach allows users to gain a deeper understanding of the landscape of proteoforms encoded by the selected gene (T2).

*Data Export*: The user may download the protein sequences encoded by all canonical and non-canonical haplotypes of the gene in the FASTA format. Additionally, the details of all the non-canonical haplotypes are available for download as a tab-separated text file, corresponding to the format produced by the *ProHap* pipeline.

### 4.2.2 Alignment of Identified Peptides

After selecting a transcript in the first section, the second section of the *Detail View* is shown, representing the protein and peptides (L3, L4). This section provides a detailed visualization of how peptides align with the selected protein, offering insights into the proteomic data associated with the gene of interest (T3).

The primary visualization in this section (Figure 4) is divided into two parts. The top part represents the canonical proteoform encoded by the selected transcript and becomes available immediately after the transcript selection. This part shows the alignment of peptides with the canonical protein sequence, providing a baseline for comparison. The mirrored bottom part of the visualization shows peptides aligned with the protein haplotype sequence and becomes available after the selection of a non-canonical proteoform in the data table of the first section. This dual representation allows users to compare the canonical and variant protein sequences side by side.

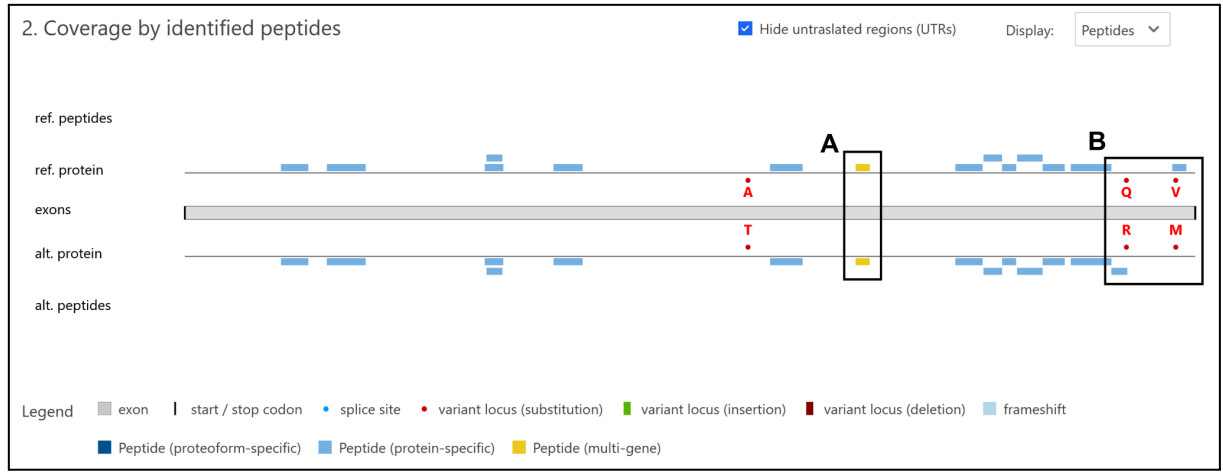

**Figure 4**: Primary visualization in the second section of the Detail view, displaying the CPN2 protein with the reference sequence at the top, and a selected non-canonical haplotype at the bottom. A: One canonical peptide matches the protein product of multiple genes. B: Only the alternative allele has been observed in a peptide for the first variant, while only the reference allele has been observed for the second variant.

In a similar way to displaying the splicing of genes, it is common to display features within a protein sequence along a linear axis representing the amino-acid position from start to end (L3). This representation is used by *UniProt* and *neXtProt* and is well understood among experts. Alternative options, such as space-filling or circular layouts, have their own advantages [49]. Space-filling layouts can show features in greater detail by stretching out the axis and providing more room, but this comes at the cost of rapid interpretability. Circular layouts can easily display relationships between distant elements along the protein; however, these relationships are already provided in the data table of the first section, and a circular layout would distort the display of quantities along the vertical axis. Therefore, we chose to conform to the state of the art and provide a linear visualization that is easy to interpret.

By default, each peptide is represented by a rectangle of uniform height, with its position and width encoding the location of the peptide within the protein. Alternatively, users may choose to view the number of spectra matched to peptides at any given position along the protein sequence. In this mode, segments of peptide sequences are represented as bars, with the height encoding the number of matched spectra and the position and width encoding the location of the segment within the protein. This flexible visualization option helps users to understand the coverage of peptides across the protein sequence.

The color of each peptide rectangle encodes the classification of the peptide based on its ability to distinguish between protein sequences. Multi-gene peptides map to the products of different genes, protein-specific peptides map to multiple sequences that are all products of the same gene, and proteoform-specific peptides map uniquely to a single proteoform sequence [4]. This color-coding allows users to quickly identify peptides that are unique to specific proteoforms or shared among multiple proteins.

The data table in the second section lists all the peptides displayed in the main figure. Each row specifies the peptide sequence, its position within the protein, the classification of the peptide, and the distribution of two quality measures of the peptide-spectrum matches (PSMs): the posterior error probability (PEP) and the difference between predicted and observed retention time. Additionally, the user can view the raw spectral data in the PRIDE database using the universal spectrum identifier (USI) [36]. The last column of the table contains links to PRIDE using the USI of the best, second-best, and median PSM, ordered by the PEP. The user may download the data table in a text format, with each row representing a peptide-spectrum match, with the quality measures and the USI specified.

*Interactions*: Users can view the sequence of the peptides by hovering the mouse over the respective elements in the visualization. Additionally, hovering the mouse over the blank spaces displays a corresponding short segment of the protein sequence. Amino acid variants in the protein sequence are highlighted in red. This feature provides immediate access to sequence information, making it easier for users to interpret the alignment and coverage of peptides within the protein.

*Data Export*: Details for all the PSMs displayed in the data table are available for download, letting the user make additional queries, or use the data in other workflows.

### 4.2.3 Highlighting by tissue, phenotype, or proteomic study

To the left of the two sections of the *Detail View*, a horizontal bar chart displays the number of elements per tissue, phenotype, or proteomic study we reanalyzed. Users select the category to display, as well as whether to display the number of PSMs, unique peptides, or the number of samples where at least one of the displayed peptides was confidently matched. Additionally, each row in the bar chart features a checkbox, allowing users to highlight the relevant category within the elements displayed in the *Protein* section of the *Detail View*. This interactive feature helps users quickly identify and focus on specific tissues, phenotypes, or proteomic studies of interest (T4).

Highlighting elements of interest, as opposed to filtering, ensures that users can see the full context of all peptide identifications, while still emphasizing the selected categories. Although highlighting multiple categories may lead to overplotting in some cases, such as with long proteins that have many identified peptides, it allows users to maintain a comprehensive view of the data. Therefore, this design choice balances the need for detailed inspection with the importance of contextual awareness.

## 5. Implementation and Code Availability

The dataset for *ProHap Explorer* is stored in a graph database implemented in neo4j, ensuring efficient querying and management of data entries and relationships between them. A Python-based pipeline is available to transform data from the standard output of *ProHap*, *ProHap Peptide Annotator* [5], and *Percolator* to [46] the *neo4j* database. The code can be accessed at [https://github.com/ProGenNo/ProHap_Graph]. The server is implemented in Python using the *Flask* library, which queries the *neo4j* database using the *cypher-shell* language, providing a backend for the application.

The user interface of *ProHap Explorer* is developed using the Svelte framework, with D3.js employed to render the visualization components. The source code for the web application, along with information about the current deployment, is available at [https://github.com/ProGenNo/ProHapExplorer]. Temporarily, ProHap Explorer is accessible with a limited dataset showing peptide-spectrum matches from the reprocessing of one proteomic study. This temporary deployment provides users with an opportunity to explore the tool and its capabilities while we continue to refine and expand the dataset.

## 6. Case Studies

To illustrate the utility of *ProHap Explorer*, we present two case studies demonstrating its application in proteogenomic analysis. These examples highlight how the tool enables researchers to investigate the impact of haplotypes on protein sequences and identify non-canonical peptides in mass spectrometry datasets.

### 6.1 Common Haplotypes of CPN2

In our previous study [5], we identified several variant peptides while searching mass spectrometry datasets against a database of protein haplotype sequences created by ProHap. Notably the CPN2 protein commonly carries a substitution of the 508th amino acid from glutamine to arginine (p.508:Q>R), encoded by the genetic variant located at position 194,341,177 on the 3rd chromosome. We have seen the variant peptide mapping to the corresponding location in the CPN2 protein in the majority of the samples processed in [5].

We have investigated the *CPN2* gene in the *Detail View* of *ProHap Explorer* (Figure 3, Figure 5). We observed two splicing alternatives present for this gene, both encoding an identical protein sequence as they only differ in the untranslated regions. Moreover, peptides encoded by the alternative allele of the substitution were identified in the underlying dataset (Figure 3A). The p.508:Q>R variant appears in three unique protein haplotypes, either in combination with one or two other amino acid substitutions (p.304:A>T, p.535:V>M), or by itself (Figure 3B). We found that the haplotype encoding only the individual amino acid substitution is more common than the others, appearing in over 44% of the participants of the 1000 Genomes Project.

Upon selecting a haplotype (p.304:A>T, p.508:Q>R, p.535:V>M), we observed that the three variants causing the amino acid substitution are linked with several combinations of variants that do not alter the protein sequence. Most commonly, they are found in combination with one variant in the transcript 3' untranslated region and two synonymous variants in the protein-coding region (Figure 5A). In the second section of the *Detail View*, we inspected the identified peptides encoded by the reference and the alternative haplotype. We noted that no peptides were matched to the reference or alternative allele of the first variant (Figure 5B). Only peptides encoded by the alternative allele of the second variant and the reference allele of the third variant were identified (Figure 5C).

Highlighting the peptides per tissue (Figure 5D) revealed that the variant peptide was uniquely observed in brain tissue from the current underlying datasets, while the peptide mapping to the reference allele at position 535 was uniquely observed in adrenal gland tissue (Figure 5C). This suggests that in the processed samples, only the alternative allele is expressed for the substitution at position 508, and only the reference allele is expressed for the substitution at position 535, consistent with our previous findings. We could not determine whether the reference or the alternative allele is expressed for the substitution at position 304, as this part of the protein was not covered by peptides at all.

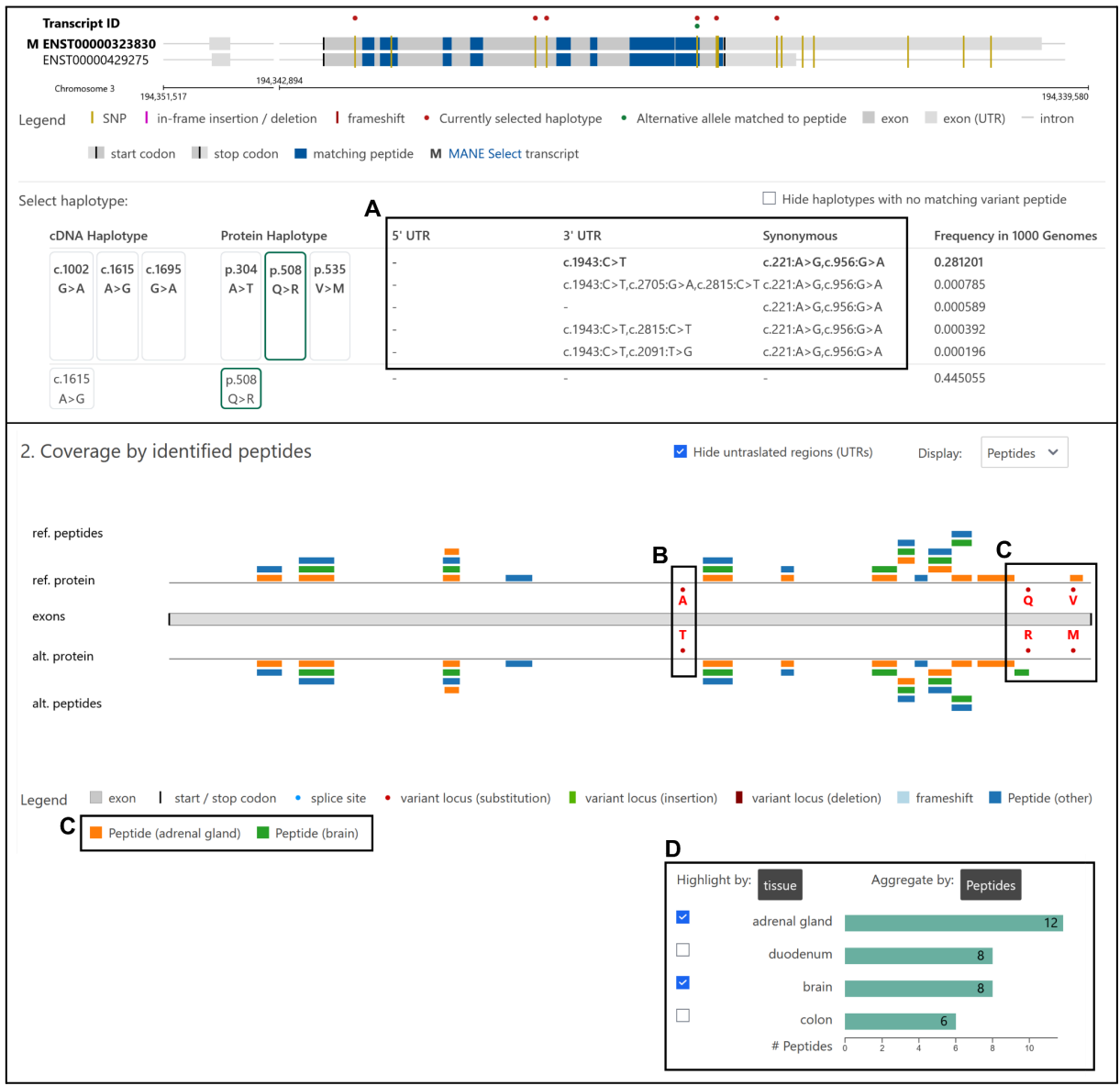

**Figure 5**: The Detail view of the *CPN2* gene after the selection of a transcript and haplotype, and highlighting peptides by tissue in the second section of the view. A: Combinations of variants which do not alter the protein sequence, but are linked with the three variants causing amino acid substitutions in the selected haplotype. B: No reference or alternative peptides have been identified. C: The peptide encoded by the alternative allele of the second variant has been uniquely identified in the brain tissue samples, while the peptide encoded by the reference allele of the third variant has been identified uniquely in the adrenal gland tissue samples. D: The categories highlighted can be selected in the visualization adjacent to the Detail view.

## 6.2 New reference sequence for CAP1

Similarly to the previous example, we have previously shown that a combination of alleles encodes five amino acid substitutions in a single peptide from the CAP1 protein [5]. This peptide was identified in a dataset of mass spectra from stem cells from a healthy donor. We previously searched these spectra against a database of protein haplotypes created by ProHap using the personal genotype of the donor, which is publicly available.

Investigating the *CAP1* gene in the *Detail View* of *ProHap Explorer* (Figure 6), we observed that a large number of splicing alternatives have been annotated for this gene. However, all the identified peptides match the alternative marked as *MANE Select*, which is the canonical splicing validated by NCBI and Ensembl [50]. Upon selecting the *MANE Select* transcript, we found that a single non-canonical haplotype has been observed for this transcript among all the participants of the 1000 Genomes panel (Figure 6B). This haplotype encodes six amino acid substitutions, all of which have been observed in at least one confidently identified peptide.

In the second section of the *Detail View*, we visualized the number of spectra matched to peptides along the protein sequence. We observed that relatively few spectra support the presence of the peptide spanning the five amino acid substitutions (Figure 6C), while a higher number of spectra support the presence of a downstream peptide, covering the sixth substitution. Upon inspecting the variant peptides in the data table, we noted that the peptide-spectrum matches have low posterior error probability, and the difference between predicted and observed chromatographic retention times is on the lower end, suggesting these are confident matches. Experts experienced in the interpretation of raw mass spectrometry data may further review these matches by clicking on one of the links in the last column of the table, navigating to the website of the PRIDE database and visualizing the spectrum annotated with names of fragment ions.

This use case demonstrates the evidence that the reference sequence of the CAP1 protein is not expected to be present in any individual, and that we can experimentally verify the presence of the variant peptides encoded by the alternative haplotype. Moreover, five of the six amino acid substitutions can only be observed in peptides if they are considered as a haplotype, highlighting the importance of inspecting haplotypes in proteins in their completeness, rather than each variant individually.

## 7. User Feedback

We conducted user interviews with four domain experts who were not involved in the development of the tool. Two of the participants are post-doc researchers, and two are PhD candidates working in the fields of proteomics and molecular biology. One of the post-docs and one of the PhD students are familiar with bioinformatics, while the remaining two participants mainly work with laboratory methods. This group provided valuable feedback on the usability and functionality of *ProHap Explorer*.

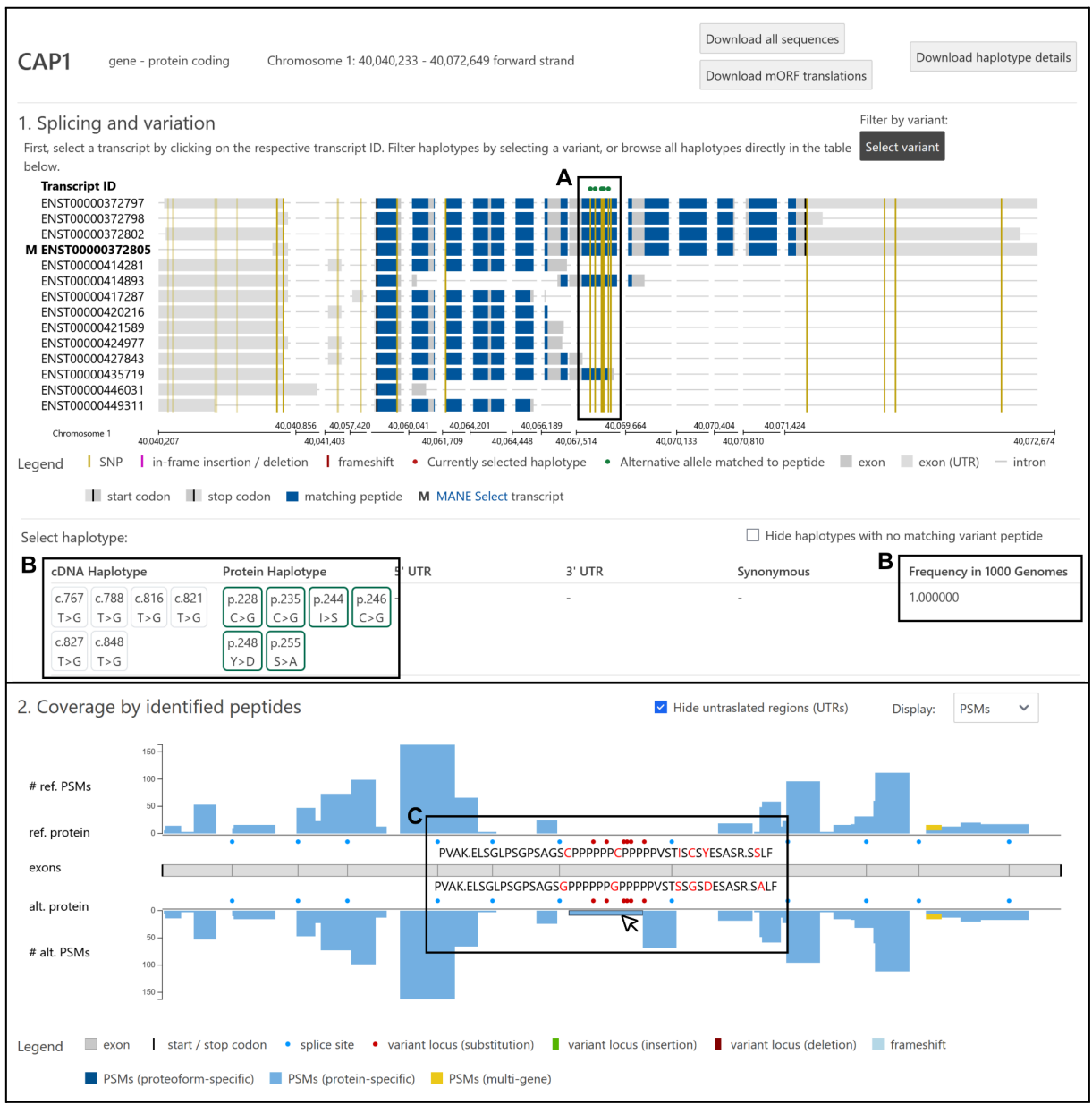

**Figure 6:** The *Detail View* of the *CAP1* gene. A: A cluster of variants affects one particular region of the gene. B: These variants all form a single haplotype, observed in all participants of the 1000 Genomes panel. C: A peptide encoded by the alternative allele of five variants has been matched to 9 different spectra. Hovering the mouse cursor over the respective element reveals the sequence of the peptide, and the corresponding section of the canonical protein.

Overall, the participants expressed satisfaction with the clarity of the interface. The linear representations of splicing alternatives and features along the protein sequence were familiar to them, making them easy to interpret. The two participants familiar with bioinformatic tools easily understood the interaction patterns and intuitively selected relevant genes, transcripts, and their haplotypes. They also clearly understood the concept of protein haplotypes and appreciated how this information could enhance their research.

The other participants took more time to familiarize themselves with the interface, as they were more accustomed to static figures. However, they all found it useful to view which parts of proteins had been identified in the mass spectrometry dataset we processed, and mentioned that they would compare this with their own data. This will allow them to interrogate whether the lack of peptides in their results might be due to overlooking the influence of genetic variation. Despite the initial learning curve, they recognized the value of the tool in providing a comprehensive view of proteogenomic data.

A few suggested additional features were noted during the interviews. Participants expressed interest in better integration with *UniProt*, in addition to the data built upon the current implementation of *ProHap*, which relies on resources from *Ensembl*. They also suggested displaying functional domains of the protein along the sequence to quickly discern which parts of the protein are of particular interest. One participant mentioned an interest in comparing different genes, although this was not a priority for the others. These suggestions highlight potential areas for future development to further enhance the utility of *ProHap Explorer* for a broader range of users.

8. **Discussion and Limitations**

   While *ProHap Explorer* offers a robust and innovative platform for visualizing proteogenomic data, there are some limitations remaining to be addressed to enhance its functionality further. Here, we discuss these limitations and outline potential solutions to ensure the tool remains effective and user-friendly.

   The usage of a graph-based backend and effective algorithms for the alignment of features in the visualizations result in a short response time when interacting with *ProHap Explorer*. This ensures an uninterrupted user experience. However, as we add more mass spectrometry data, particularly when thousands of spectra match to each peptide, we anticipate an increase in the latency between user interactions and the rendering of components. If this problem arises, we will need to further optimize the algorithms or pre-compute some of the values that are currently calculated online. These adjustments are feasible and will ensure that the tool remains responsive for large datasets.

   Integration with *UniProt* and the inclusion of extended textual descriptions of proteins will help experts find their proteins of interest more easily. While this feature is not yet implemented, it is an addition that will significantly enhance the usability of *ProHap Explorer*. By providing more context and detailed information about each protein, users will be able to navigate the data more efficiently and make more informed decisions.

In some cases, the visualizations can become cluttered, making it difficult to focus on specific features of interest. A zoom feature would help address this issue by allowing users to hone in on particular areas of the visualization. Although this feature is not currently available, the existing modes of interaction and the ability to download raw data enable users to inspect each gene in great detail. This ensures that, despite some visual clutter, users can still access and analyze the information they need. Future updates will aim to incorporate zoom functionality to further improve the user experience.

## 9. Conclusion

We have presented *ProHap Explorer*, a novel visual exploration tool designed to interrogate the influence of common haplotypes on the human proteome. With the recent introduction of *ProHap*, experts can now create customized sequence databases of protein haplotypes to search mass spectrometry data, or use one of the databases published alongside *ProHap*. However, a resource that would allow experts to browse the haplotypes and their influence on protein sequences using public mass spectrometry data was missing. *ProHap Explorer* bridges this gap by providing an interactive visual interface, enabling experts to explore this novel resource effectively.

Our user interviews with participants of varying levels of expertise in proteomics confirmed the utility of *ProHap Explorer*. Participants appreciated the opportunity to investigate whether haplotypes have any consequences on their proteins of interest and whether any non-canonical peptide sequences have been identified in public mass spectrometry datasets. The familiar linear representations and interactive elements facilitated easy navigation and interpretation of the data. Additionally, our interviews highlighted a further need to integrate other resources, such as *UniProt*, to add new layers of information on protein function. This will inform our future steps in the development of *ProHap Explorer*.

**Acknowledgements**

This work was supported by the Research Council of Norway (project #301178 to MV), Stiftelsen Trond Mohn Foundation (Mohn Center for Diabetes Precision Medicine), and the University of Bergen. LK was supported by the Wallenberg AI, Autonomous Systems and Software Program (WASP) funded by the Knut and Alice Wallenberg Foundation. The computations on publicly-available data were performed on the Norwegian Research and Education Cloud (NREC), using resources provided by the University of Bergen and the University of Oslo. https://www.nrec.no.

This research was funded, in whole or in part, by the Research Council of Norway 301178. A CC BY or equivalent license is applied to any Author Accepted Manuscript (AAM) version arising from this submission, in accordance with the grant's open access conditions.## References

[1]  S. McCarthy et al., "A reference panel of 64,976 haplotypes for genotype imputation," *Nat Genet*, vol. 48, no. 10, Art. no. 10, Oct. 2016, doi: 10.1038/ng.3643.

[2]  A. I. Nesvizhskii, "Proteogenomics: concepts, applications and computational strategies," *Nat Methods*, vol. 11, no. 11, Art. no. 11, Nov. 2014, doi: 10.1038/nmeth.3144.

[3]  G. Menschaert and D. Fenyö, "Proteogenomics from a bioinformatics angle: A growing field," *Mass Spectrometry Reviews*, vol. 36, no. 5, pp. 584–599, 2017, doi: 10.1002/mas.21483.

[4]  J. Vašíček et al., "Finding haplotypic signatures in proteins," *GigaScience*, vol. 12, p. giad093, Jan. 2023, doi: 10.1093/gigascience/giad093.

[5]  J. Vašíček et al., "ProHap enables human proteomic database generation accounting for population diversity," *Nat Methods*, pp. 1–5, Dec. 2024, doi: 10.1038/s41592-024-02506-0.

[6]  A. Auton et al., "A global reference for human genetic variation," *Nature*, vol. 526, no. 7571, Art. no. 7571, Oct. 2015, doi: 10.1038/nature15393.

[7]  J. Vasicek, "Protein haplotype sequences obtained by ProHap from the 1000 Genomes Project data set." Zenodo, Jul. 08, 2024. doi: 10.5281/zenodo.12671237.

[8]  D. Wang et al., "A deep proteome and transcriptome abundance atlas of 29 healthy human tissues," *Molecular Systems Biology*, vol. 15, no. 2, p. e8503, Feb. 2019, doi: 10.15252/msb.20188503.

[9]  E. de Hoffmann, "Tandem mass spectrometry: A primer," *Journal of Mass Spectrometry*, vol. 31, no. 2, pp. 129–137, 1996, doi: 10.1002/(SICI)1096-9888(199602)31:2<129::AID-JMS305>3.0.CO;2-T.

[10]  K. Verheggen, H. Raeder, F. S. Berven, L. Martens, H. Barsnes, and M. Vaudel, "Anatomy and evolution of database search engines-a central component of mass spectrometry based proteomic workflows," *Mass Spectrom Rev*, vol. 39, no. 3, pp. 292–306, May 2020, doi: 10.1002/mas.21543.

[11]  D. Fenyö and R. C. Beavis, "A Method for Assessing the Statistical Significance of Mass Spectrometry-Based Protein Identifications Using General Scoring Schemes," *Anal. Chem.*, vol. 75, no. 4, pp. 768–774, Feb. 2003, doi: 10.1021/ac0258709.

[12]  A. T. Kong, F. V. Leprevost, D. M. Avtonomov, D. Mellacheruvu, and A. I. Nesvizhskii, "MSFragger: ultrafast and comprehensive peptide identification in mass spectrometry–based proteomics," *Nat Methods*, vol. 14, no. 5, pp. 513–520, May 2017, doi: 10.1038/nmeth.4256.

[13]  L. Käll, J. D. Storey, and W. S. Noble, "Non-parametric estimation of posterior error probabilities associated with peptides identified by tandem mass spectrometry," *Bioinformatics*, vol. 24, no. 16, pp. i42–i48, Aug. 2008, doi: 10.1093/bioinformatics/btn294.

[14]  A. Declercq et al., "MS2Rescore: Data-driven rescoring dramatically boosts immunopeptide identification rates," *Molecular & Cellular Proteomics*, p. 100266, Jul. 2022, doi: 10.1016/j.mcpro.2022.100266.

[15]  H. M. Umer et al., "Generation of ENSEMBL-based proteogenomics databases boosts the identification of non-canonical peptides," *Bioinformatics*, vol. 38, no. 5, pp. 1470–1472, Mar. 2022, doi: 10.1093/bioinformatics/btab838.

[16]  X. Cao and J. Xing, "PrecisionProDB: improving the proteomics performance for precision medicine," *Bioinformatics*, vol. 37, no. 19, pp. 3361–3363, Oct. 2021, doi: